\documentclass{article}
\pdfoutput=1
\usepackage{amsmath}
\usepackage{amssymb}
\usepackage{graphicx}
\usepackage{color}
\usepackage[usenames,dvipsnames,svgnames,table]{xcolor}
\usepackage{soul}
\usepackage{subfig}
\usepackage{ulem}
\usepackage{float}
\usepackage{multirow}

\pdfpageheight\paperheight
\pdfpagewidth\paperwidth

\usepackage{jcappub}

\definecolor{darkred}{RGB}{175,0,0}

\newcommand{\BACCUS}{\texttt{BACCUS}}

\newcommand{\beq}{\begin{equation}}
\newcommand{\btheta}{\boldsymbol{\theta}}
\newcommand{\bSigma}{\boldsymbol{\Sigma}}
\newcommand{\bDelta}{\boldsymbol{\Delta}}

\newcommand{\post}{\mathcal{P}}
\newcommand{\eeq}{\end{equation}}

\usepackage{array}
\def\ga{ \mathrel{\rlap{\raise 0.59ex
 \hbox{$>$}}{\lower 0.59ex \hbox{$\sim$}}}}
\def\kmsmpc{\,{\rm km}\,{\rm s}^{-1}{\rm Mpc}^{-1}}

\def\bigstrut{\vrule width0pt height1.3em depth0.7em}

\begin{document}
\title{Conservative cosmology: combining data with allowance for unknown systematics}
\author[a,b]{Jos\' e Luis Bernal}
\author[c]{John A. Peacock}

\affiliation[a]{ICC, University of Barcelona, IEEC-UB, Mart\' i i Franqu\` es, 1, E08028
Barcelona, Spain}
\affiliation[b]{Dept. de F\' isica Qu\` antica i Astrof\' isica, Universitat de Barcelona, Mart\' i i Franqu\` es 1, E08028 Barcelona,
Spain}
\affiliation[c]{Institute for Astronomy, University of Edinburgh, Royal Observatory, Blackford Hill, Edinburgh,
EH9 3HJ, UK}

\emailAdd{joseluis.bernal@icc.ub.edu}
\emailAdd{jap@roe.ac.uk}

\abstract{When combining data sets to perform parameter inference, the results
will be unreliable if there are unknown systematics in data or models.
Here we introduce a flexible methodology, $\BACCUS$: BAyesian
Conservative Constraints and Unknown Systematics, which deals in a
conservative way with the problem of data combination, for any degree
of tension between experiments. We introduce parameters that
 describe a bias in each model parameter for each class of experiments.
A conservative posterior for the model parameters is then obtained by
marginalization both over these unknown shifts and over the width of
their prior.  We contrast this approach with an existing 
 method in which each individual likelihood is scaled,
comparing the performance of each approach and their combination in
application to some idealized models. Using only these rescaling 
 is not a suitable approach for the current
observational situation, in which internal null tests of the errors
are passed, and yet different experiments prefer models that are in
poor agreement. The possible existence of large shift systematics
cannot be constrained with a small number of data sets, leading to
extended tails on the conservative posterior distributions.  We
illustrate our method with the case of the $H_0$ tension between
results from the cosmic distance ladder and physical measurements that
rely on the standard cosmological model.
}

\maketitle

\hypersetup{pageanchor=true}

\section{Introduction}\label{sec:Introduction}

For two decades or more, the standard $\Lambda$-Cold Dark Matter
($\Lambda$CDM) cosmological model has succeeded astonishingly well in
matching new astronomical observations, and its parameters are
precisely constrained
(e.g. \cite{Planckparameterspaper,Alam_bossdr12,Betoule14_jla}). However,
more recent work has persistently revealed tensions between high and
low redshift observables. In the case of the Hubble constant, $H_0$,
the best direct measurement using cepheids and supernovae type Ia
\cite{RiessH0_2016} is in $3.4\sigma$ tension with the value inferred
assuming $\Lambda$CDM using Planck observations
\cite{Planckparameterspaper}. Planck and weak lensing surveys both
measure the normalization of density fluctuations via the combination
$\Omega_m^{0.5}\sigma_8$, and their estimates were claimed to be in
$2.3\sigma$ tension by the KiDS collaboration \cite{Hildebrandt_kids}
-- although recent results from DES are less discrepant
\cite{DES_Y1_cluslens}. 
These inconsistencies are not currently definitive
(e.g. \cite{Heavens_evidence}), but they raise the concern that
something could be missing in our current cosmological understanding.
It could be that the $\Lambda$CDM model needs extending, but it could
also be that the existing experimental results suffer from
unaccounted-for systematics or underestimated errors. 

When inconsistent data are combined naively, it is well understood
that the results risk being inaccurate and that formal errors may be
unrealistically small.
  For this reason,
much emphasis is placed on tests that can be used to assess the
consistency between two data sets (e.g.  
\cite{Marshall_robustness,  Verde_tension2d, Seehars_surprise}). 
 We refer the interested reader
to \cite{Charnock_tension, Lin_ioi} for more methodologies but also
for a comprehensive comparison between different measures of
discordance.  Another approach is the posterior predictive
distribution, which is the sampling distribution for new data given
existing data and a model, as used in e.g., \cite{Feeney_PPD}.
However, these methods are not really helpful in cases of mild
tension, where a subjective binary decision is required as to whether
or not a genuine inconsistency exists. 

Unknown systematics can be modelled as the combination of two distinct
types. Type 1 systematics affect the random scatter in the
measurements (and therefore the size of the errors in a parameterised
model), but do not change the maximum-posterior values for the
parameters of the model. In contrast, type 2 systematics offset the
best-fitting parameters without altering the random errors; they are
completely equivalent to shifts in the parameters of the model without
modifying the shape of the posterior of each parameter. While the former are commonly detectable through internal evidence, the latter are more dangerous and they can only reveal themselves when independent experiments are compared. Much of our discussion will
focus on this class of systematic. 
With a
  detailed understanding of a given experiment, one could do better
  than this simple classification; but here we are trying to capture
  `unknown unknowns' that have evaded the existing modelling of
  systematics, and so the focus must be on the general character of
  these additional systematics. 

Taking all this into account, there is a need for a general
conservative approach to the combination of data. This method should
allow for possible unknown systematics of both kinds and it should
permit the combination of data sets in tension with an agnostic
perspective. Such a method will inevitably yield uncertainties in the
inferred parameters that are larger than in the conventional approach,
but having realistic uncertainties is important if we are to establish
any credible claims for the detection of new physics.

The desired method can be built using a hierarchical approach. Hierarchical schemes have been used widely in cosmology, e.g. 
 to model in more detail the dependence of the
parameters on each measurement in the case of $H_0$ and the cosmic
distance ladder \cite{Feeney_H0}, or the cosmic shear power spectrum~ \cite{Alsing_bhm_lens}.  While 
the extra parameters often model
  physical quantities, our application simply requires empirical
  nuisance parameters.  The introduction of extra parameters 
   to deal with
  data combination was first introduced in the pioneering discussion
  of \cite{Press_hyperpars}. A more general formulation was provided
  by \cite{Lahav_hp} and refined in \cite{Hobson_hp} (H02
  hereinafter).  This work assigns a free weight to each data set,
  rescaling the logarithm of each individual likelihood 
  (which is equivalent to rescaling the errors of each
  experiment if the likelihood is Gaussian), in order to achieve an
  overall reduced $\chi^2$ close to unity.  The H02 method yields
  meaningful constraints when combining data sets affected by type 1
  errors, and it detects the presence of the errors by comparing the
  relative evidences of the conventional combination of data and their approach.   However, this method is not appropriate for
obtaining reliable constraints in the presence of type 2 systematics,
where we might find several experiments that all have reduced $\chi^2$
values of unity, but with respect to different best-fitting
models. 
  H02
do not make our distinction between different types of systematics,
but in fact they do show an example where one of the data sets has a
systematic type 2 shift (see Figures 3 \& 4 of H02). Although their method does detect the presence of the systematic, we do not feel that it gives a satisfactory posterior in this case,
  for reasons discussed below in section \ref{sec:contrast}.

Here we present a method called $\BACCUS$\footnote{A python
    package implementing is publicly available
    in \href{https://github.com/jl-bernal/BACCUS}{https://github.com/jl-bernal/BACCUS}.}, BAyesian
Conservative Constraints and Unknown Systematics, which is designed to
deal with systematics of both types. Rather than weighting each data
set, the optimal way to account for type 2 systematics is to consider
the possibility that the parameters preferred by each experiment
  are offset from the true values. Therefore, extra parameters shift the model parameters when computing each
individual likelihood, and marginalized posteriors of the model
parameters will account for the possible existence of systematics in a
consistent way. Moreover, studying the marginalized posteriors of 
these new parameters can reveal which experiments are most strongly
affected by systematics. 

This paper is structured as follows. In Section \ref{sec:Methodology},
we introduce our method and its key underlying assumptions.  In
Section \ref{sec:examples}, we consider a number of illustrative
examples of data sets constructed to exhibit both concordance and
discordance, contrasting the results from our approach with those of
H02.  We then apply our method to a genuine
cosmological problem, the tension in $H_0$, in Section
\ref{sec:H0}. Finally, a summary and some general discussion of the
results can be found in Section \ref{sec:Conclusions}. We use the
Monte Carlo sampler \texttt{emcee} \cite{emcee} in all the cases where
Monte Carlo Markov Chains are employed.

\section{Overview of assumptions and methodology}\label{sec:Methodology}

We begin by listing the key assumptions that underlie our statistical
approach to the problem of unknown systematics. Firstly, we will group all codependent experiments in different \textit{classes}, and consider each of them independent from the others. For
example, observations performed with the same telescope or analyzed
employing the same pipeline or model assumptions will be considered in
the same class, since all the really dangerous systematics would be in
common. Then, our fundamental assumption regarding systematics will be that 
 a experiment belonging to each of these classes is equally likely to commit an error of a given
magnitude, and that these errors will be randomly and independently
drawn from some prior distribution.   
 Attempts have been made to allow for dependence
between data sets when introducing scaling parameters 
 as in H02
(see \cite{Ma_hpcorr}). We believe that a similar extension of our
approach should be possible, but we will not pursue this complication
here.

With this preamble, we can now present the formalism to be used.
Consider a model $M$, parameterised by a set of model parameters,
$\btheta$ (we will refer to $M(\btheta)$ as $\btheta$ for simplicity), which are to be constrained by several data sets, $\bf{D}$. The corresponding posterior, $\post\left(\btheta\lvert\bf{D}\right)$, and the likelihood, $\post\left(\bf{D}\lvert\btheta\right)$, are related by the Bayes theorem:
\begin{equation}
\post\left(\btheta\lvert\bf{D}\right) = \frac{\post\left(\btheta\right)\post\left(\bf{D}\lvert\btheta\right)}{\post\left(\bf{D}\right)},
\end{equation} 
where $P(\btheta)$ is the prior. We
will consider flat priors in what follows and concentrate on the
likelihood, unless otherwise stated. 
For parameter inference, we can
drop the normalization 
 without loss of
generality.

We can account for the presence of the two types of systematics in the
data by introducing new parameters. 
  For type 2
systematics, the best fit values of the parameters for each
experiment, $\tilde{\btheta}_i$, may be offset from the true value by
some amount. We introduce a shift parameter, 
$\Delta_{\theta_j}^i$, for each parameter $\theta_j$ and class $i$ of
experiments.  For type 1 systematics, we follow H02 and introduce a
 rescaling parameter $\alpha_i$ 
 which weights the logarithm of the likelihood
of each class of experiments; if the likelihood is Gaussian, this is equivalent to rescaling each individual $\chi^2$ and therefore the covariance.  
 Considering $n_i$ data points for the class of experiments $i$, the likelihood is now:
\begin{equation}
\post(\btheta,\boldsymbol{\alpha},\lbrace\boldsymbol{\Delta}_{\btheta}\rbrace\lvert{\bf{D}}) 
\propto \prod_i \alpha_i^{n_i/2}\, \exp\left[-\frac{\alpha_i}{2}\left(\chi^2_{{\rm bf},i} + \Delta\chi^2_i(\btheta+\bDelta_{\btheta}^i-\tilde{\btheta}_i) \right)\right],
\label{eq:lkl1D}
\end{equation}
where $\chi^2_{\rm bf}$ is the minimum $\chi^2$, corresponding to the
best-fit value of ${\btheta}$, $\tilde{ \boldsymbol{\theta} }$. 
Here, we use the notation
$\lbrace\boldsymbol{\Delta}_{\btheta}\rbrace$ to indicate the
vector of shift parameters for each parameter of the
model. There is a different vector of this sort for every class of
experiments, indexed by $i$.

For rescaling parameters, 
 H02 argue that the prior should be taken
as: 
\begin{equation}
\post(\alpha_i) = \exp[-\alpha_i]
\label{eq:prior_alpha}
\end{equation}
so that the mean value of $\alpha_i$ over the prior is unity,
i.e. experiments estimate the size of their random errors correctly on
average.  One might quarrel with this and suspect that underestimation
of errors could be more common, but we will retain the H02 choice;
this does not affect the results significantly. In realistic cases
where the number of degrees of freedom is large and null tests are
passed so that $\chi^2_{{\rm bf},i}\simeq n$, the scope for rescaling
the errors will be small and $\alpha_i$ will be forced to be close to
unity.  For the prior on shift parameters, we choose a zero-mean
Gaussian with a different unknown standard deviation determined by
$\sigma_{\theta^j}$, corresponding to each parameter $\theta_j$ and
common to all classes of experiments.  Furthermore, it is easy to imagine
systematics that might shift several parameters in a correlated way,
so that the prior on the shifts should involve a full covariance
matrix, $\bSigma_{\Delta}$, containing the variances,
$\sigma_{\theta^j}^2$, in the diagonal and off-diagonal terms obtained
with the correlations $\rho_{j_1,j_2}$ for each pair of shifts
($\Sigma_{j_1,j_2}=\rho_{j_1,j_2}\sigma_{\theta^{j_1}}\sigma_{\theta^{j_2}}$). Thus
the assumed prior on the shifts is:
\begin{equation}
\post(\lbrace\boldsymbol{\Delta}_{\btheta}\rbrace\lvert\boldsymbol{\sigma}_{\theta},\boldsymbol{\rho}) \propto \prod_i^N \lvert\bSigma_\Delta\lvert^{-1/2}\exp\left[-\frac{1}{2} {\bDelta_{\theta}^i}^{\rm T}\boldsymbol{\Sigma}^{-1}_\Delta {\bDelta_{\theta}^i}\right].
\label{eq:priorDelta1D}
\end{equation}

We now need to specify the hyperpriors for the covariance matrix
  of the shifts, $\bSigma_{\Delta}$. Our philosophy here is to seek an
  uninformative hyperprior: it is safer to allow the data to limit the
  degree of possible systematic shifts, rather than imposing a
  constraining prior that risks forcing the shifts to be
  unrealistically small. 
  
Different options of priors for covariance matrices are discussed by e.g. \cite{Alvarez_covmathp}. In order to ensure independence among variances and correlations, we use a separation strategy, applying different priors to variances and correlations (e.g. \cite{Barnard_sepstrat}).  A covariance matrix
  can be expressed as $\bSigma =
  \boldsymbol{S}\boldsymbol{R}\boldsymbol{S}$, with $\boldsymbol{S}$
  being a diagonal matrix with $S_{jj}=\sigma_{\theta^j}$ and
  $\boldsymbol{R}$, the correlation matrix, with $R_{ii} = 1$ and
  $R_{ij} = \rho_{ij}$. As hyperprior for each of the covariances we choose a lognormal distribution ($\log \sigma=N(b,\xi)$, where $N(b,\xi)$ is a Gaussian distribution in $\log\sigma$ with mean value $b$ and variance $\xi$). In the case of the correlation matrix, we use the LKJ distribution \cite{Lewandowski_LKJ} as hyperprior, which  depends only on the parameter $\eta$: for $\eta=1$, it is an uniform
  prior over all correlation matrices of a given order; for $\eta>1$,
  lower absolute correlations are favoured (and vice versa for
  $\eta<1$). The parameters $b$, $\xi$ and $\eta$ can be chosen to
  suit the needs of the specific problem. We prefer to be as agnostic
  as possible, so we will choose $\eta = 1$ and $b$ and $\xi$ such as the hyperprior of each covariance is broad enough to not to force the shifts to be small.  

The final posterior can be marginalized over all added parameters, leaving the conservative distribution of the model 
parameters $\btheta$, that is the main aim of this work.  This
immediately provides a striking insight: a single experiment gives no
information whatsoever. 
 It is
only when we have several experiments that the possibility of large
$\boldsymbol{\sigma}_{\btheta}$ starts to become constrained (so that
the $\lbrace\boldsymbol{\Delta}_{\btheta}\rbrace$ cannot be too
large). In the case of consistent data, as the shifts are drawn from a Gaussian distribution, only small shifts are favoured (as the individual likelihoods would not overlap otherwise). 
If, on the
other hand, only two data sets are available and there is a tension
between them regarding some parameter $\theta_j$, the prior width
$\sigma_{\theta^j}$ could be of the order of such tension, but much
larger values will be disfavoured. 

However, an alternative would be 
 to obtain the marginalized posteriors
of $\lbrace\boldsymbol{\Delta}_{\btheta}\rbrace$. This tells us the
likely range of shifts that each data set needs for each parameter, so
that unusually discrepant experiments can be identified by the
system. As we will see in examples below, this automatically results
in their contribution to the final posterior being downweighted. If
one class of experiments has shifts that are far beyond all others,
this might give an objective reason to repeat the analysis without it,
but generally we prefer not to take this approach: judging whether an
offset is significant enough to merit exclusion is a somewhat
arbitrary decision, and is complicated in a multidimensional parameter
space. Our formalism automatically downweights data sets as their
degree of inconsistency grows, and this seems sufficient.

\section{Application to illustrative examples}\label{sec:examples}

\subsection{Shift parameters in the one-parameter Gaussian case}\label{sec:Gaussian}
In order to exhibit all the features of our method more clearly, we
first apply the formalism to the simple model in which there is only
one parameter ($\btheta = a$) and the probability density functions
(PDFs) of $a$ for the $N$ individual experiments are Gaussian. In this
case, 
we can rewrite Equation
\ref{eq:lkl1D} as:
\begin{equation}
\post(a,\boldsymbol{\alpha},\boldsymbol{\Delta}_a, \sigma_a\lvert{\bf{D}}) \propto \prod_i^N  \alpha_i^{n_i/2}\exp\left[-\frac{1}{2}\alpha_i\left(\chi^2_{i,\rm
    bf}+\Delta\chi_i^2(a+\Delta_a^i)\right)\right].
\end{equation}
We apply the prior for rescaling parameters (Equation \ref{eq:prior_alpha})  
and marginalize over each $\alpha_i$ to obtain the marginalized
posterior for a single class of experiments:
\begin{equation}
\post_i(a,\Delta_a^i,\sigma_a\lvert D_i) \propto \left(\Delta\chi_i^2(a+\Delta_a^i)+2\right)^{-(n_i/2+1)}.
\label{eq:MarginAlpha}
\end{equation}
For large $n_i$, $\chi_i^2+2\simeq \chi_i^2$, and the
  right hand side of Equation \ref{eq:MarginAlpha} is proportional to 
  \smash{$\exp[-(\Delta\chi_i^2/2)(n_i/\chi^2_{i,\rm bf})]$}, which in
  effect instructs us to rescale parameter uncertainties according to
  \smash{$(\chi^2_{\nu,i})^{1/2}$}, where \smash{$\chi^2_{\nu,i}$} is the reduced
  \smash{$\chi_i^2$} for the class of experiments $i$. But it can be
  assumed that experiments will pursue internal null tests to the
  point where $\chi^2_\nu\simeq 1$; thus in practice rescaling parameters can do little to erase tensions.

Assuming hereafter that experimenters will achieve $\chi_\nu = 1$
  exactly, we can now focus on the novel feature of our approach,
  which is the introduction of shift parameters. Then, the posterior can be written as
\begin{equation}
\post(a,\bDelta_a,\sigma_a\lvert {\bf D}) \propto  
\prod_i^N \sigma_a^{-1}\exp\left[-\frac{1}{2}\sum_{k=1}^{n_i} \left(\frac{\left(y_i^k(a+\Delta_a^i) - D_i^k\right)^2}{{\sigma_i^k}^2}\right) + \frac{{\Delta_a^i}^2}{2\sigma_a^2}\right],
\end{equation}
where $y_i^k(x)$ is the theoretical prediction to fit to the
measurement $D_i^k$ of the class of experiments $i$, with error
$\sigma_i^k$. Note that the width of the prior for the shifts, $\sigma_a$, is the same
for all data sets, by assumption.  Marginalizing over the shifts,
 then the posterior of each class of experiments is:
\begin{equation}
\post_i(a,\sigma_a\lvert{\bf D}) \propto \prod_{k=1}^{n_i} ({\sigma_i^k}^2 + \sigma_a^2)^{-1/2}
\exp\left[ -{1\over 2} \frac{\left(y_i^k(a)-D_i^k\right)^2}{\left({\sigma_i^k}^2+\sigma_a^2\right)^2}\right].
\label{eq:Margin1D}
\end{equation}
Therefore, our method applied to a model with only one parameter and
Gaussian likelihoods reduces to the  convolution of the original
  posteriors with a Gaussian of width $\sigma_a$.

\begin{figure}[t]
\centering
\includegraphics[width=0.9\textwidth]{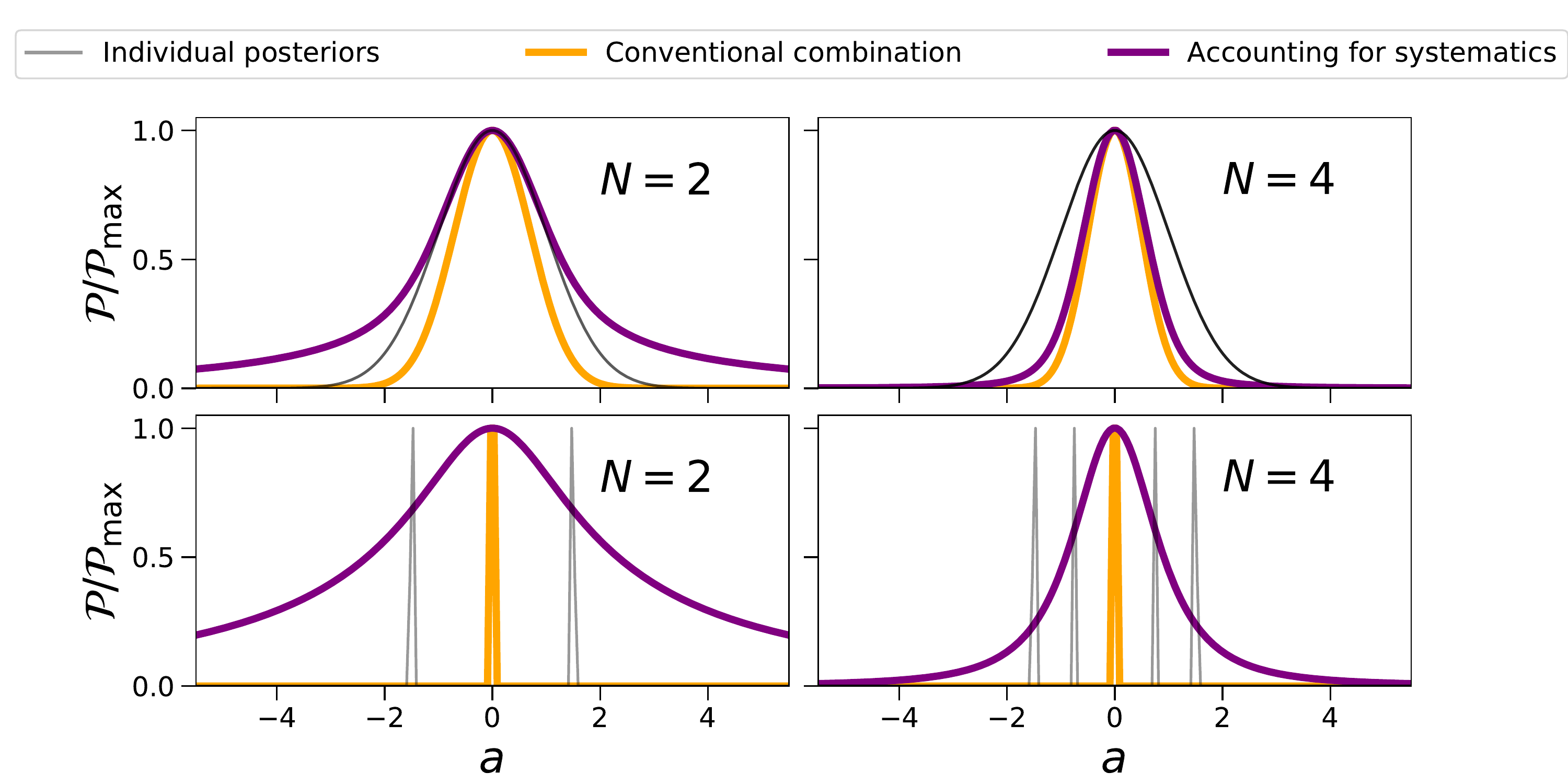}
\caption{Comparison of the results obtained using shift parameters 
 and the
  conventional approach to combining data sets in a model with only one
  parameter, $a$, and 
  $N$ data sets whose individual posteriors 
  are Gaussians. We show individual posteriors in black, the
  posteriors obtained with the conventional approach in orange and the
  posterior obtained with our approach, in purple. The dependence of
  the posterior on the number of data sets for the exactly consistent case is
  shown in the top panels, while strongly inconsistent cases are shown in the
  bottom panels.}
\label{fig:1D_2-4}
\end{figure}

Finally, we need to marginalize over $\sigma_a$.  
   Consider for example a
  hyperprior wide enough to be approximated as uniform in $\sigma_a$,
  and suppose that all the $N$ data sets agree on $\tilde{a}_i=0$ and
  all the errors, $\sigma_i^k = \sigma_i$, are identical. Then we can
  derive the marginalized posterior in the limit of small and large
  $a$: 
\begin{equation}
 \post(a\lvert{\bf D}) \propto \left\{
  \begin{array}{@{}ll@{}}
    1-\exp\left[\frac{(N-1)a^2}{2\sigma_i^2}\right], & \text{for}\ a\ll 1 \\
    a^{1-N}, & \text{for}\ a\gg 1
  \end{array}\right.
\end{equation}
For values of $a$ close to $\tilde{a}_i$ the posterior presents a
Gaussian core, whose width is $\sigma_i/\sqrt{N-1}$, in contrast with
the conventional $\sigma_i/\sqrt{N}$ from averaging compatible
data. 
  For values of $a$ very far from $\tilde{a}_i$, the
posterior has non-Gaussian power-law tails. For $N=2$ these are so
severe that the distribution cannot be normalized, so in fact three
measurements is the minimum requirement to obtain well-defined
posteriors. As will be discussed in Section \ref{sec:Conclusions}, one
can avoid such divergence by choosing harder priors on $\Delta_a^i$ or
$\sigma_a$, but we prefer to be as agnostic as possible. Nonetheless, these `fat tails' on the posterior are less of an issue as $N$
increases. These two
aspects of compatible data can be appreciated in the top panels of
Figure \ref{fig:1D_2-4}.  The message here is relatively optimistic:
provided we have a number of compatible data sets, the conservative
posterior is not greatly different from the conventional one. 

Alternatively, we can consider an example of strongly incompatible
data. Let the $N$ data sets have negligible $\sigma_i$ and suppose the
corresponding $\tilde{a}_i$ are disposed symmetrically about $a=0$
with spacing $\epsilon$, e.g. $\tilde{a} = (-\epsilon,0,+\epsilon)$
for $N=3$.  This gives a marginalized posterior that depends on
$N$. For example, the tails follow a power law: $\post(a\lvert {\bf
  D})~\propto~ (1+~4a^2/\epsilon^2)^{-1/2}$ for $N=2$, $\post~(a\lvert {\bf
  D})~\propto~ (1+3a^2/2\epsilon^2)^{-1}$ for $N=3$, etc., with an
asymptotic dependence of $(a/\epsilon)^{1-N}$ for $N\gg 1$. So, as in
the previous case, the posterior cannot be normalized if $N=2$, but it
rapidly tends to a Gaussian for large $N$. This case is shown in the
bottom panels of Figure \ref{fig:1D_2-4}.
  The appearance of these extended tails on the posterior
  is a characteristic result of our method, and seems inevitable if
  one is unwilling in advance to limit the size of possible shift
  systematics. The power-law form depends in detail on the hyperprior,
  but if we altered this by some power of $\sigma_a$, the result would
  be a different power-law form for the `fat tails' still with the generic non-Gaussinianity.

\begin{figure}[t]
\centering
\includegraphics[width=0.9\textwidth]{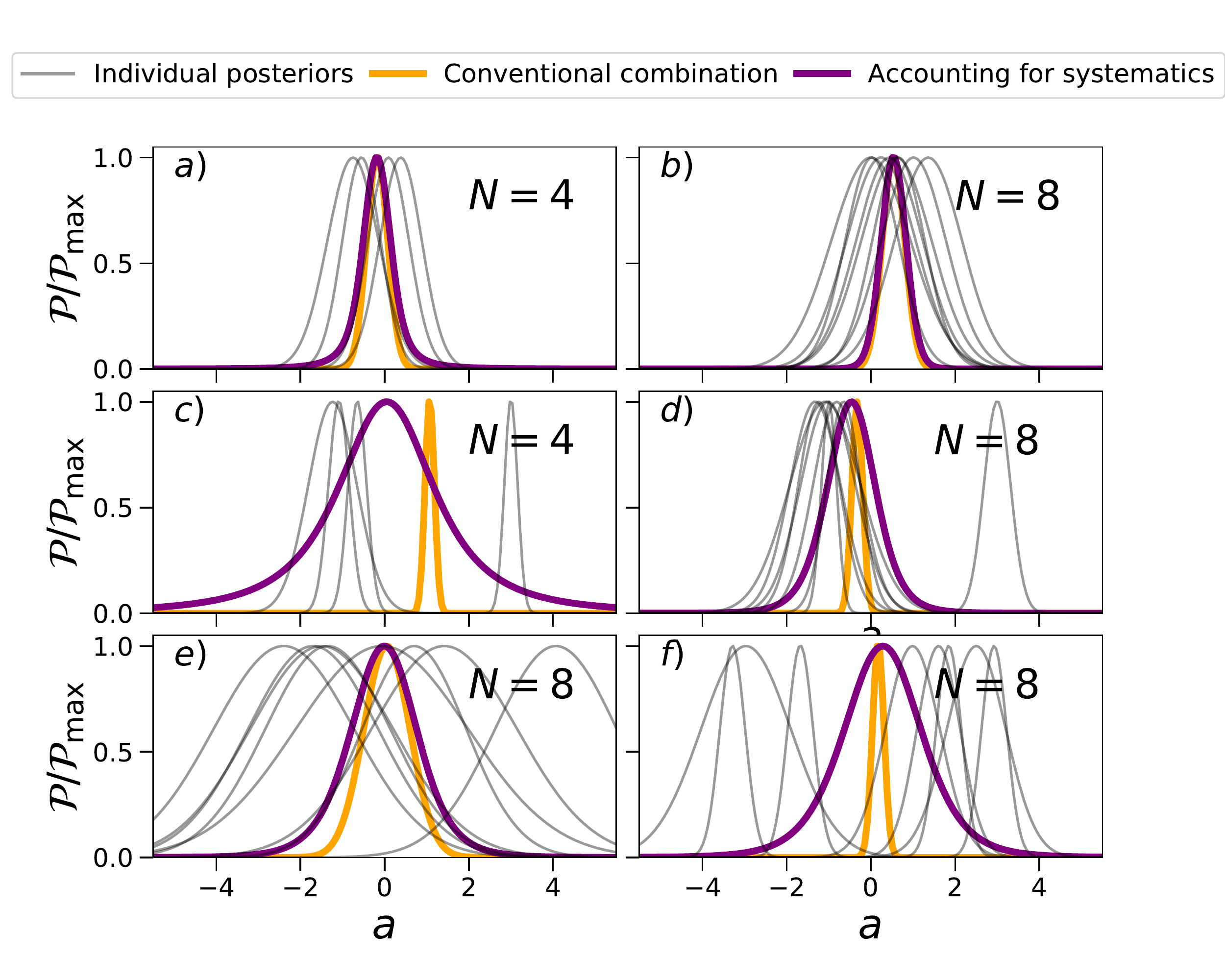}
\caption{The same as Figure \ref{fig:1D_2-4}, but considering cases in
  which all the data sets are consistent (panels $a$ and $b$), only
  one is discrepant with the rest (panels $c$ and $d$), eight data
  sets with scatter larger than the errors (panel $e$) and eight data sets with random
  values of the best fit and errors (panel $f$).}
\label{fig:1D_general}
\end{figure}

We also show in Figure \ref{fig:1D_general} some more realistic
examples, starting with mock consistent data that are drawn from a
Gaussian using the assumed errors (rather than $\tilde a=0$), but then
forcing one or more of these measurements to be discrepant.  As with
the simple $\tilde a=0$ example, we see that the results for several
consistent data sets approach the conventional analysis for larger $N$
(panels $a$ and $b$). But when there is a single discrepant data set,
the posterior is much broader than in the conventional case (panel
$c$).  Nevertheless, as the number of consistent data sets increases,
the posterior shrinks to the point where it is only modestly broader
than the conventional distribution, and where the single outlying
measurement is clearly identified as discrepant (panel $d$). Thus our
prior on the shifts, in which all measurements are assumed equally
likely to be in error, does not prevent the identification of a case
where there is a single rogue measurement. However, these examples do
emphasize the desirability of having as many distinct classes of
measurement as possible, even though this may mean resorting to
measurements where the individual uncertainties are larger.  
 Additional coarse information
 can play an important role in limiting the tails on the posterior,
especially in cases where there are discordant data sets (see panel
$d$).  Finally, we also show examples where the scatter of the
individual best-fit is larger than the individual uncertainties of the
data sets, so the size of the shifts are larger and our posterior is
broader than the one obtained with conventional approach (panel $e$),
and a case with several inconsistent measurements (with best-fit and
errors distributed randomly), for which our posterior is centred close
to 0 with a width set by the empirical distribution of the data (panel
$f$).

\subsection{Contrasting shift and rescaling parameters} \label{sec:contrast}
If we ignore the constraints on $\alpha_i$ and consider only the
relative likelihoods (with width of the distribution determined by
$\sigma_i$), then there is an illuminating parallel between the
effects of rescaling and shift parameters.
 Compare Equation \ref{eq:Margin1D}, where all $\alpha$ have been already marginalized over ($\post_1$), with
H02's method ($\post_2$):
\begin{equation}
\post_1 \propto \prod_i (\sigma_a^2+\sigma_i^2)^{-1/2}\exp\left[-\frac{1}{2}\sum_i \frac{(a-\tilde{a}_i)^2}{\sigma_a^2+\sigma_i^2}  \right]\, ; \quad\quad
\post_2 \propto \prod_i \alpha_i^{1/2}\sigma_i^{-1}\exp\left[-\frac{1}{2}\sum_i \frac{\alpha_i(a-\tilde{a}_i)^2}{\sigma_i^2}  \right],
\label{eq:compH02}
\end{equation}
these two expressions are clearly the same if
$\alpha_i=\left(1+\sigma_a^2/\sigma_i^2 \right)^{-1}$. However, there
is a critical difference: while there is an $\alpha_i$ for each class
of experiments, we only consider a single $\sigma_a$, which
participates in the prior for all the shift parameters of all classes
of experiments. 
 
 On the other hand, if different $\sigma_{\theta^j}$ for each
  data sets were to be used, this would be equivalent to a double use
  of rescaling parameters.  
    Furthermore, in the case of having
several experiments with inconsistent results, the posterior using
only rescaling parameters would be a multimodal distribution peaked at
the points corresponding to the individual posteriors, as seen in Figures 3 \& 4 of H02.
  We feel that this is not a
  satisfactory outcome: it seems dangerously optimistic to believe
  that one out of a flawed set of experiments can be perfect when
  there is evidence that the majority of experiments are
  incorrect. Our aim should be to set conservative constraints, in
  which all experiments have to demonstrate empirically that they are
  {\it not\/} flawed (i.e. `guilty until proved innocent').

\subsection{Examples with multiple parameters}\label{sec:Comp}
  The approach to models with multiple
parameters differs conceptually from the one-parameter case: there are
several families of shifts, $\lbrace\bDelta_\theta\rbrace$,   with their
corresponding covariance matrix. 
  A convenient simple illustration 
   is provided
by the example chosen by H02: consider data sets sampled from
different straight lines. Thus, the model under consideration is $y =
mx + c$, where $y$ \& $x$ are the information given by the data and
$m$ \& $c$ are the parameters to constrain.

\begin{figure}[h!]
\centering
\includegraphics[width=0.95\textwidth]{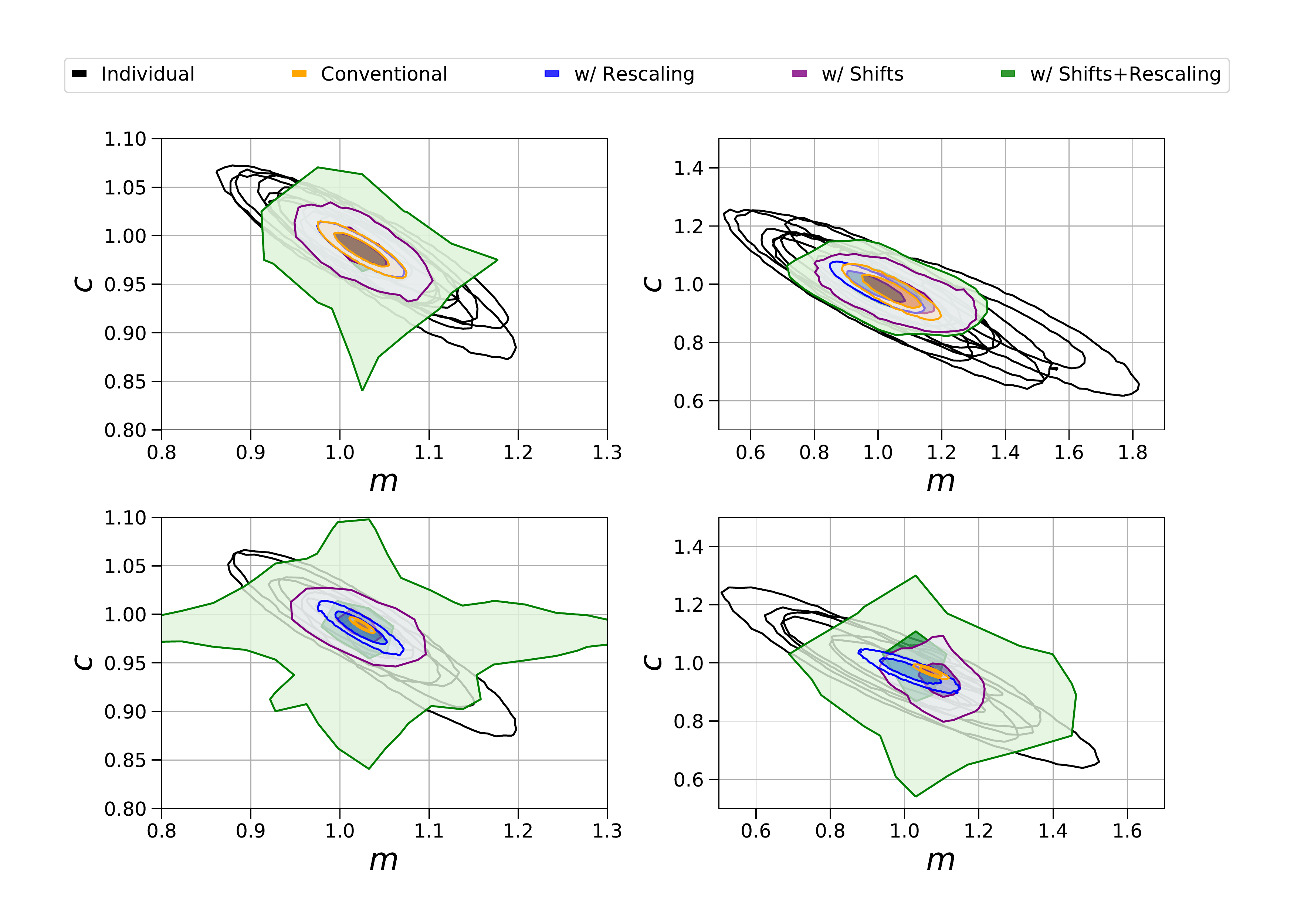}
\caption{Constraints for six data sets sampled from a straight line
  with slope $m=1$ and intercept $c=1$ ($\lbrace D_1\rbrace$ and
  $\lbrace D_2\rbrace$). We show the individual posteriors in black
  and the results from using the conventional approach in orange, using
  rescaling parameters, 
   in blue, using shift parameters, in
  purple, and using both 
   in green. Top left:
  all data sets have 50 points. Top right: all data sets have 5
  points. Bottom panels: as in the top panels, but the errors of
  $\lbrace D_2\rbrace$ are underestimated a factor 5.}
\label{fig:comp_cons}
\end{figure}

\begin{figure}[h!]
\centering
\includegraphics[width=0.95\textwidth]{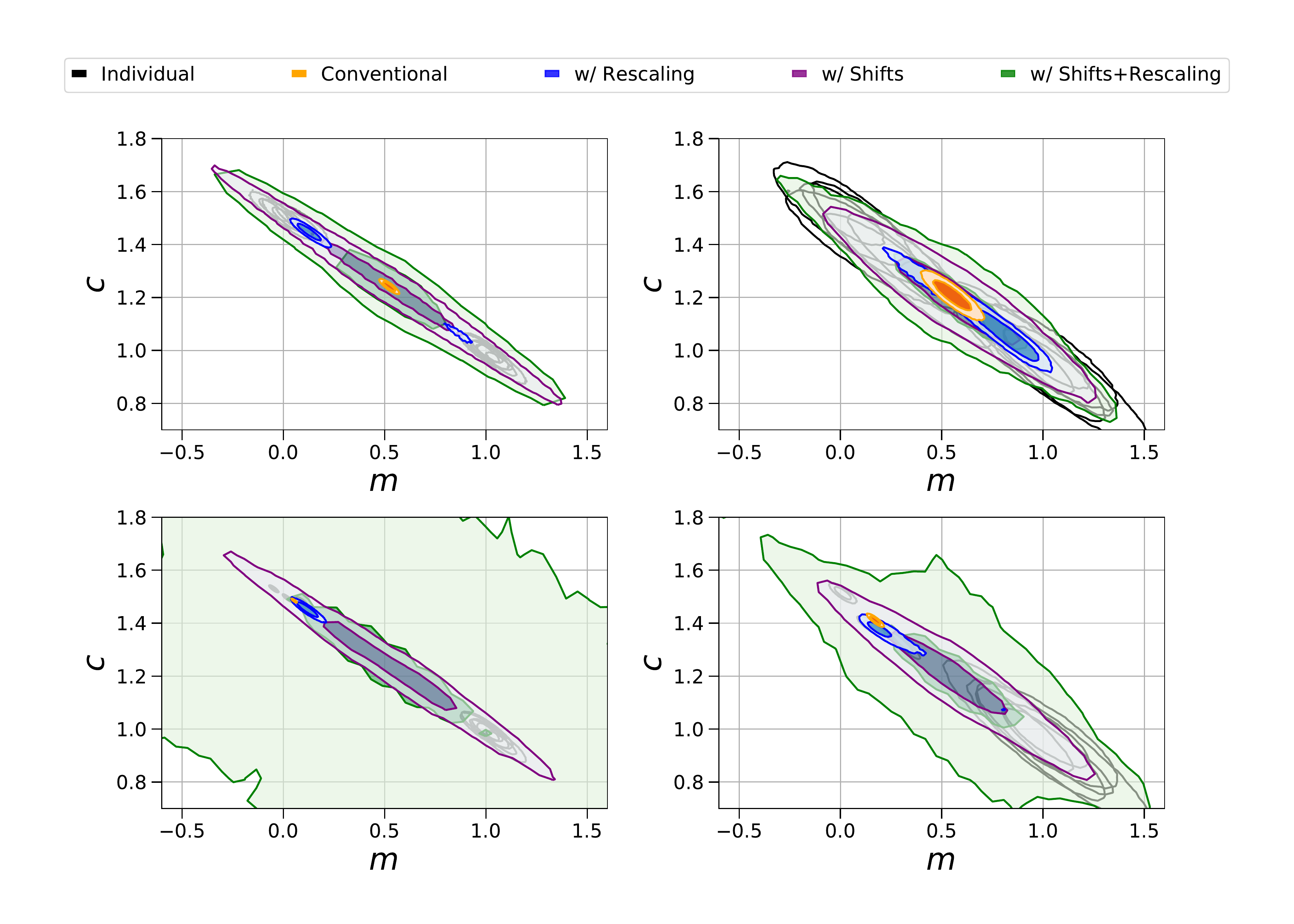}
\caption{As Figure \ref{fig:comp_cons} but using $\lbrace D_3\rbrace$
  (with slope $m=0$ and intercept $c = 1.5$) instead of $\lbrace
  D_2\rbrace$.}
\label{fig:comp_incons1}
\end{figure}

\begin{figure}[h!]
\centering
\includegraphics[width=0.95\textwidth]{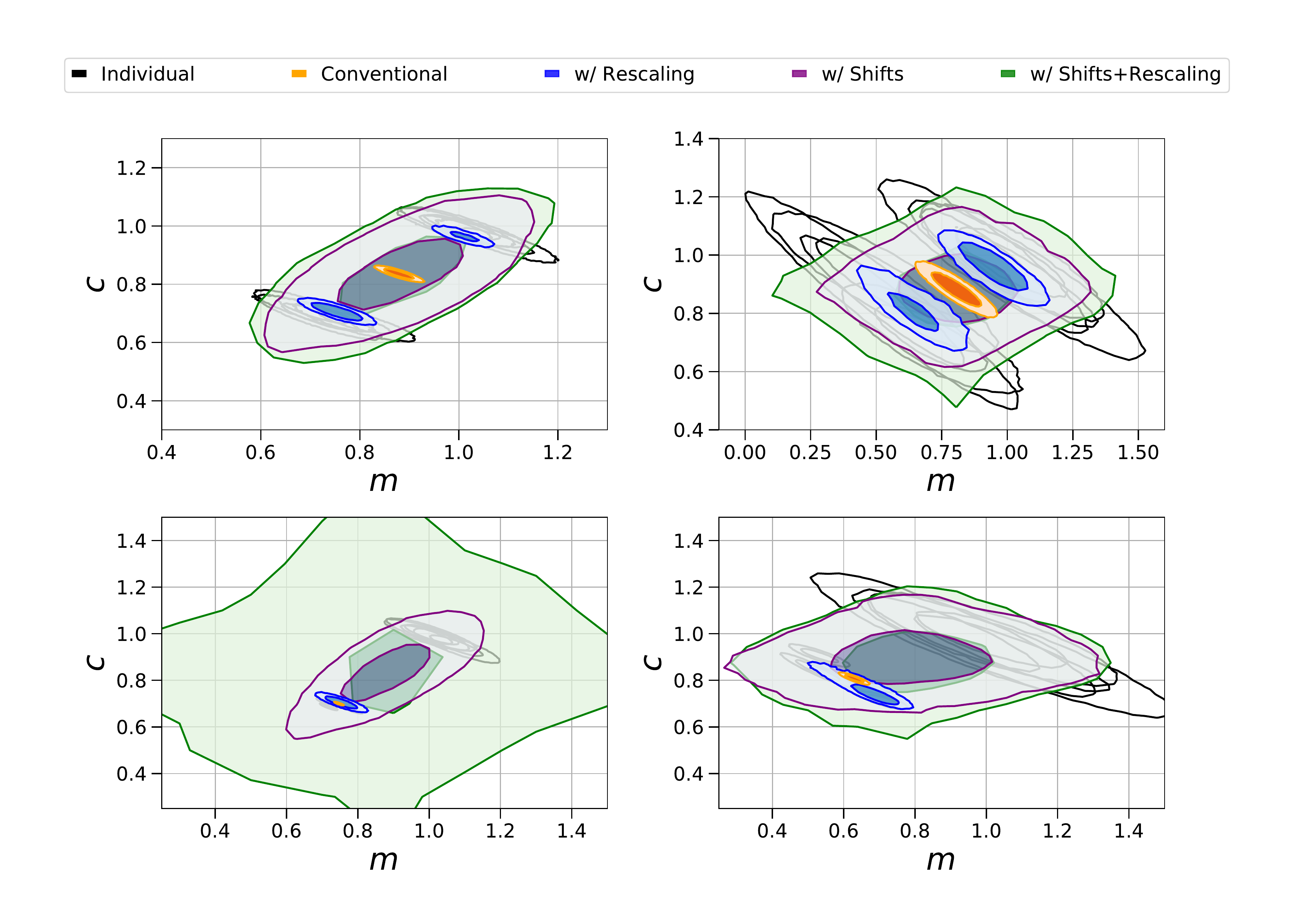}
\caption{As Figure \ref{fig:comp_cons} but using $\lbrace D_4\rbrace$
  (with slope $m=0.7$ and intercept $c = 0.7$) instead of $\lbrace
  D_2\rbrace$.}
\label{fig:comp_incons2}
\end{figure}

We consider three different straight lines for which we sample the
data, $D_i$: $\lbrace D_1\rbrace$ and $\lbrace
D_2\rbrace$~$\equiv$~$\lbrace m= c = 1\rbrace$; $\lbrace
D_3\rbrace\equiv \lbrace m= 0,\,c = 1.5\rbrace$; and $\lbrace
D_4\rbrace\equiv\lbrace m= c = 0.7 \rbrace$. For all $D_i$, we
consider three independent data sets (so $N=6$ when combining i.e.,
$D_1$ and $D_2$) and assume $\sigma_y = 0.1$ for every data point.  We
combine $\lbrace D_1\rbrace$ with $\lbrace D_2\rbrace$ in Figure
\ref{fig:comp_cons}, with $\lbrace D_3\rbrace$ in Figure
\ref{fig:comp_incons1}, and with $\lbrace D_4\rbrace$ in Figure
\ref{fig:comp_incons2}. Note the change of scale in each panel. In all
cases, we study four situations corresponding to the combination of:
all data sets with 50 or 5 points and errors correctly estimated or
underestimated by a factor 5 (only in data sets from $\lbrace
D_2\rbrace$, $\lbrace D_3\rbrace$ or $\lbrace D_4\rbrace$).  We
  use lognormal priors with $b=-2$ and $\xi=16$ 
   both for $\sigma_m$
  and $\sigma_c$, and a LKJ distribution with $\eta = 1$ 
   as the shifts hyperprior.  We
  show the individual posteriors of each data set in black; the
  results using the conventional approach in orange; the constraints
  using only rescaling parameters 
   in blue; using only shift parameters
   in purple; and using both 
  in green. The occasional noisy shape of the latter is due to the numerical
complexity of sampling the parameter space using rescaling and shifts.
Generally, in this case the uncertainties are somewhat larger than in
the case of using only shifts, except when individual errors are
poorly estimated and the credible regions are much larger. This is
because rescaling parameters gain a large weight in the analysis in
order to recover a sensible $\chi_{\nu,i}^ 2$, which permits shifts that are too
large for the corresponding prior (given that the corresponding
likelihood is downweighted by small values of $\alpha_i$). This can be seen
comparing green and purple contours in the bottom panels of Figures \ref{fig:comp_cons}, \ref{fig:comp_incons1} \&
\ref{fig:comp_incons2}.

As can be seen in Figure \ref{fig:comp_cons}, if the data sets are
consistent and the errors are correctly estimated (top left panel),
rescaling parameters 
 have rather little effect on the final
posterior.  This supports our argument in Equation
\ref{eq:MarginAlpha} and below. 
 On the other hand, when errors are underestimated
(bottom left panel), the recovered posterior is similar to the one in
which the errors are correctly estimated. When the data sets contain
smaller number of points (right panels) the results are qualitatively
similar. 

As expected, rescaling parameters 
yield conservative
constraints accounting for type 1 systematics, but it 
is not a good choice if the data sets are
not consistent. As shown in Figures \ref{fig:comp_incons1} \&
\ref{fig:comp_incons2}, the posterior for this case is multimodal,
implying that the true values for the parameters are equally likely to
correspond to one of the reported sets of values and ruling out values
in between experiments, as 
 foreshadowed in the previous section. 
  Moreover, when the data sets
are inconsistent and the errors of some of them underestimated, the
constraints 
 tend to favour only
the values corresponding to such data sets (although with larger
uncertainties than the conventional approach). Therefore, although
rescaling parameters help to diagnose if any data set is suffering
from both types of systematics, they cannot be used to obtain
meaningful constraints if type 2 systematics are present.

On the other hand, using shift parameters gives constraints with
  larger uncertainties, allowing values between the results of the
  individual data sets and accepting the possibility that experiments might be 
   polluted by unaccounted-for type 2 systematics 
  (as
it is the case in Figures \ref{fig:comp_incons1} \&
\ref{fig:comp_incons2}).  Surprisingly, they also provide correct
conservative constraints when only type 1 systematics are present,
with results similar to those obtained using only rescaling parameters
(see, e.g., the bottom left panel of Figure \ref{fig:comp_cons}).

\section{Applications to cosmology: $H_0$}\label{sec:H0}
In order to illustrate how our method performs in a problem of real
interest, we apply it to the tensions in $H_0$. 
This tension has been studied from different perspectives in the
literature. One of the options is to perform an independent analysis
of the measurements to check for systematics in a concrete constraint,
e.g., by including rescaling parameters to consider type 1 systematics in
each measurement used to constrain $H_0$ \cite{Cardona_H0} or using a
hierarchical analysis to model in more detail all the probability
distribution functions \cite{Feeney_H0}. Another possibility is to
consider that this tension is a hint of new physics, rather than a
product of unaccounted-for systematics, and therefore explore if other
cosmological models ease it or if model independent approaches result
in constraints that differ from the expectations of $\Lambda$CDM (see
\cite{BernalH0} and references therein).

Here we propose a third way. We consider all the existing independent
constraints of $H_0$ from low redshift observations and apply
$\BACCUS$ to combine them and obtain a conservative joint constraint
of $H_0$, accounting for any possible 
scale or shift systematic in each class of
experiments (grouped as described in Section~\ref{sec:Data}). We
use only low redshift observations in order to have
a consensus conservative constraint to confront with early Universe
constraints from CMB observations. We assume a $\Lambda$CDM background
expansion and use the cosmic distance ladder as in
\cite{Heavens:2014rja,Cuesta:2014asa,StandardQuantities}.

\subsection{Data and modelling}\label{sec:Data}
In this section we describe the data included in the analysis. As
discussed in Section~\ref{sec:Gaussian}, 
the size of the uncertainties using $\BACCUS$
are smaller for a larger number of classes of experiments, even if the
individual errors are larger. Therefore, we include all independent
constraints on $H_0$ from low redshift observations available,
independent of the size of their error bars.   In principle, we
  should use the exact posterior reported by each experiment, but
  these are not always easily available.  Therefore, we use the reported 68\% credible limits 
   in the case of the direct measurements of $H_0$, assuming a Gaussian likelihood.
  The resulting error in the tails of the posterior
  is one form of systematic, which \BACCUS\ should be able to absorb.
The different classes of experiments are grouped as described below:

\begin{itemize}
\item {\bfseries Direct measurements using the distance ladder.} We
  include as different classes of experiments direct measurements that
  use different standard candles or distance anchors. These are:
  the three independent measurements used in Riess et al. 2016 (see table 6 in \cite{RiessH0_2016}), 
  the relation between the integrated H$\beta$ line luminosity and the
  velocity dispersion of the ionized gas in HII galaxies and giant HII
  regions\cite{FernandezArenas_H0}, megamasers
  \cite{Reid_megamaser, Kuo_megamaser, Gao_megamaser} and    the
  $H_0$ value measured by the Cosmic Flows project
  \cite{CosmicFlows3}. 
   Finally, we use the
  direct measurement coming from the standard siren
  \cite{standard_siren} from the neutron star merger whose
  gravitational wave was detected by VIRGO and LIGO collaborations
  \cite{NSmerger} and whose electromagnetic counterpart was also
  detected by several telescopes \cite{NSmerger_EM}.  We do not include
  the measurement using the Tip of the Red Giant Branch from
  \cite{Jang_H0trgb} because such analysis uses anchors and
  measurements included in the analysis of Riess et al. (2016)
  \cite{RiessH0_2016}.

\item {\bfseries Baryon Acoustic Oscillations (BAO).} Assuming an
  underlying expansion history, BAO measurements constrain the low
  redshift standard ruler, $r_{\rm s}h$ (see
  e.g. \cite{StandardQuantities}), where $r_{\rm s}$ is the sound
  horizon at radiation drag and $h=H_0/100$. Measurements of the
  primordial deuterium abundance can be used to break this degeneracy
  \cite{Addison_bao13,Aubourg_2015}, given that they can be used to
  infer the physical density of baryons, $\Omega_bh^2$ \cite{Cooke_deu2017}.
   We use BAO measurements from the following
  galaxy surveys: Six Degree Field Galaxy Survey (6dF)
  \citep{Beutler11}, the Main Galaxy Sample of Data Release 7 of Sloan
  Digital Sky Survey (SDSS-MGS) \citep{Ross15}, the galaxy sample of
  Baryon Oscillation Spectroscopic Survey Data Release 12 (BOSS DR12)
  \cite{Alam_bossdr12}, the Lyman-$\alpha$ forest autocorrelation from
  BOSS DR12 \cite{Bautista_bao_lya} and their cross correlation with
  quasars \cite{Bourboux_bao_lyaqso}, the reanalysed measurements of
  WiggleZ \citep{Kazin14_wz}, and the measurement using quasars at
  $z=1.52$ \cite{Ata_eboss}. We use anisotropic measurements when
  available (including their covariance) and account for the
  covariance between the different redshift bins within the same
  survey when needed. We consider BOSS DR12 and WiggleZ measurements
  as independent because the overlap of both surveys is very small,
  hence their correlation (always below 4\%) can be neglected
  \citep{Beutler16_overlap, Cuesta16_bao}. For our analysis, we
  consider observations of different surveys or tracers (i.e., the
  autocorrelation of the Lyman-$\alpha$ forest and its cross
  correlation with quasars are subject to different systematics) as
  different classes of experiments. 

\item {\bfseries Time delay distances.} Using the time delays from the
  different images of strong lensed quasars it is possible to obtain a
  good constraint on $H_0$ by using the time delay distance 
   if an expansion history is
  assumed. We use the three measurements of the H0LiCOW project
  \cite{H0_holicow} as a single class of experiment

\item {\bfseries Cosmic clocks.} 
  Differential ages of old elliptical galaxies provide estimate of the
  inverse of the Hubble parameter, $H(z)^{-1}$ \cite{Jimenez_C}. We
  use a compilation of cosmic clocks measurements including the
  measurement of \cite{Moresco16}, which extends the prior compilation
  to include both a fine sampling at $0.38<z<0.48$ using BOSS Data
  Release 9, and the redshift range up to $z\sim2$. As all cosmic
  clock measurements have been obtained from the same group using similar analyses, we
  consider the whole compilation as a single class of experiment.

\item {\bfseries Supernovae Type Ia.} 
   As we want to focus mostly on $H_0$,
  we use the Joint Light curve Analysis (JLA) of Supernovae Type Ia
  \cite{Betoule14_jla} as a single class of experiment to constrain the unnormalized expansion history
  $E(z) = H(z)/H_0$, hence tighter constraints on the matter density
  parameter, $\Omega_M$, are obtained. 

\end{itemize}

We do not consider the assumption of a $\Lambda$CDM-like expansion history (which connects BAO, time delay distances, cosmic clocks and supernovae) as a source of systematic errors which couples different class of experiments (since it affects each observable in a different way). Therefore, we can neglect any correlation among these four probes. In order to interpret the above experiments, we need a model
  that contains three free parameters: $H_0$, $\Omega_ch^2$, and
  $\Omega_bh^2$.    $\Omega_bh^2$ will only be constrained by a prior
coming from \cite{Cooke_deu2017} and, together with $\Omega_ch^2$ and
$H_0$, allows us to compute $r_{\rm s}$ and break the degeneracy
between $H_0$ and $r_{\rm s}$ in BAO measurements. As we focus on
$H_0$ and variations in $\Omega_bh^2$ do not affect $E(z)$
significantly, we do not apply any shift to $\Omega_bh^2$.  We compute a grid of values of $100\times r_{\rm s}h$ for
different values of $H_0$, $\Omega_ch^2$ and $\Omega_bh^2$ using the
public Boltzmann code CLASS \cite{Lesgourgues:2011re,Blas:2011rf}
before running the analysis and interpolate the values at each step of
the MCMC to obtain $r_{\rm s}$ in a rapid manner\footnote{We
  make a grid for $100\times r_{\rm s}h$ in order to minimize the
  error in the interpolation ($\lesssim 0.1\%$). This grid is
  available upon request.}.

\subsection{Results}\label{sec:Results}
\begin{figure}[t]
\centering
\includegraphics[width=0.8\textwidth]{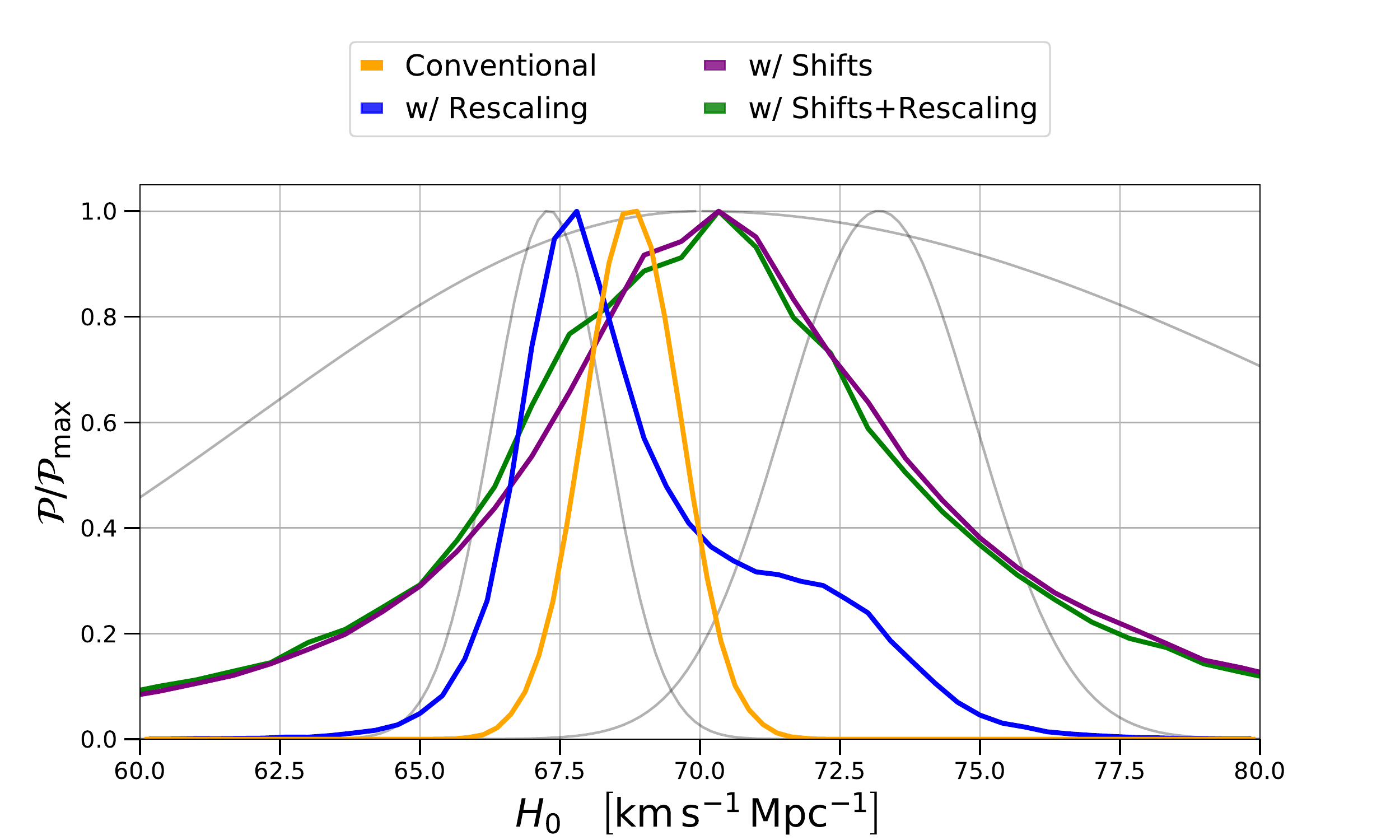}
\caption{Marginalized $H_0$ posterior distributions obtained from the
  combination of marginalized $H_0$ constraints from the local measurement of Riess et al. 2016, Planck and the neutron star merger. 
   We show
  results with the standard approach (orange), with only rescaling
   (blue), with only shifts 
    (purple) and
  with both rescaling and shifts 
   (green).}
\label{fig:H0_1D_planckr16}
\end{figure}

In this section we show the results using $\BACCUS$ when addressing
the tension in $H_0$. We compare them with the results obtained using
the conventional approach and the methodology introduced in H02. 
First, we consider marginalized measurements of $H_0$. Ideally, we would apply $\BACCUS$ to Riess et al. 2016 and Planck measurements. However, as stated in Section \ref{sec:Methodology}, this method can not be applied to only two measurements. Thus, we use the independent and much broader measurement coming from the neutron star merger \cite{standard_siren} in order to constrain the tails of the final posterior. These results can be found in Figure \ref{fig:H0_1D_planckr16}. Even with the inclusion of a third measurement, the tails of the posterior when shift parameters are added are still too large and therefore the conservative constraints are very week (due to the low number of experiments included). On the other hand, adding only rescaling parameters results in a bimodal distribution. In order to obtain relevant conservative constraints, more observations need to be included in the analysis.

 As the next step, we perform an analysis with more data and compare the results of the different methodologies to the combination of the data listed in Table \ref{tab:H0_1D}, as recently used in \cite{DES_H0}. 
Since
marginalized constraints in clear tension are combined, this is a case
where $\BACCUS$ is clearly necessary. 
  We use the lognormal hyperprior with $b=-2$ and
  $\xi=16$ for the hyperprior of variance of the shifts in both cases.

\begin{figure}[t]
\centering
\includegraphics[width=0.8\textwidth]{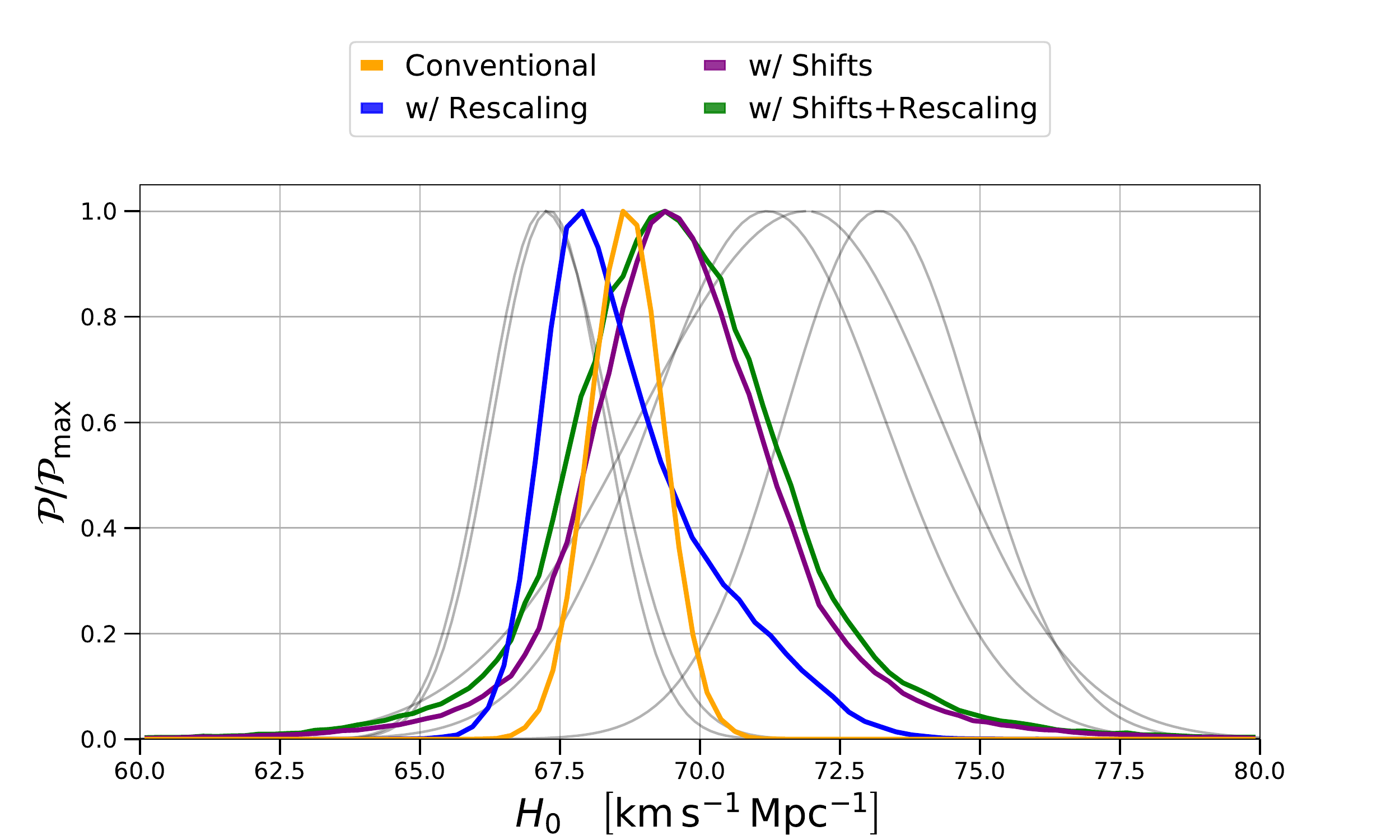}
\caption{Marginalized $H_0$ posterior distributions obtained from the
  combination of marginalized $H_0$ constraints from the experiments listed in Table \ref{tab:H0_1D}. 
   We show
  results with the standard approach (orange), with only rescaling 
   (blue), with only shifts 
    (purple) and
  with both rescaling and shifts 
   (green).}
\label{fig:H0_1D}
\end{figure}

\begin{table}[htbp]
\centering
\begin{tabular}{|c|c|l|}
\hline
	&	Experiment/Approach	&	$H_0$ ($\kmsmpc$) \\
\hline
\multirow{5}{*}{Individual Measurements}	& DES\cite{DES_H0}	&	\quad$67.2^{+1.2}_{-1.0}\bigstrut$ \\
	&	Planck \cite{Planckparameterspaper}	& \quad$67.3\pm 1.0\bigstrut$ \\
	& SPTpol \cite{SPTpol}	& \quad$71.2\pm 2.1\bigstrut$	\\	
	& H0LiCOW \cite{H0_holicow}	& \quad$71.9^{+2.4}_{-3.0}\bigstrut$	\\
	& Riess et al. 2016 \cite{RiessH0_2016}	& \quad$73.2\pm 1.7\bigstrut$ \\
\hline
\multirow{4}{*}{This work}	& Conventional combination & \ $68.7\pm 0.6(\pm 1.2)\bigstrut$ \\
	& Rescaling param.	& \ $67.8^{+1.8}_{-0.6}(^{+4.1}_{-1.3})\bigstrut$\\
	& Shift param.			& \ $69.5^{+1.7}_{-1.4}(^{+4.7}_{-3.4})\bigstrut$ \\
	& Shift + rescaling param. & \ $69.4^{+2.1}_{-1.4}(^{+4.9}_{-3.8})\bigstrut$ \\
\hline

\end{tabular}
\caption{Individual marginalized constraints on $H_0$ combined to evaluate the
  performance of our method in a real one dimensional problem. In the
  bottom part, we report highest posterior density values and 68$\%$
  (95$\%$ in parenthesis)  credible limits obtained combining the individual measurement using different
  kind of parameters.}
\label{tab:H0_1D}
\end{table}

The results of this comparison can be found in Figure \ref{fig:H0_1D}
and Table \ref{tab:H0_1D}, where we report the marginalized highest posterior
density values and 68$\%$ (95$\%$ in parenthesis) credible limits 
 and the individual measurements used. As
expected, the results using $\BACCUS$ peak among the individual best
fits and have larger uncertainties than using the conventional
approach. However, comparing with the individual constraints, the
result seems more sensible.  
 There is a small difference between the combined result
reported in~\cite{DES_H0} and our result using the conventional
approach due to using different samplers.

\begin{figure}[h!]
\centering
\includegraphics[width=0.75\textwidth]{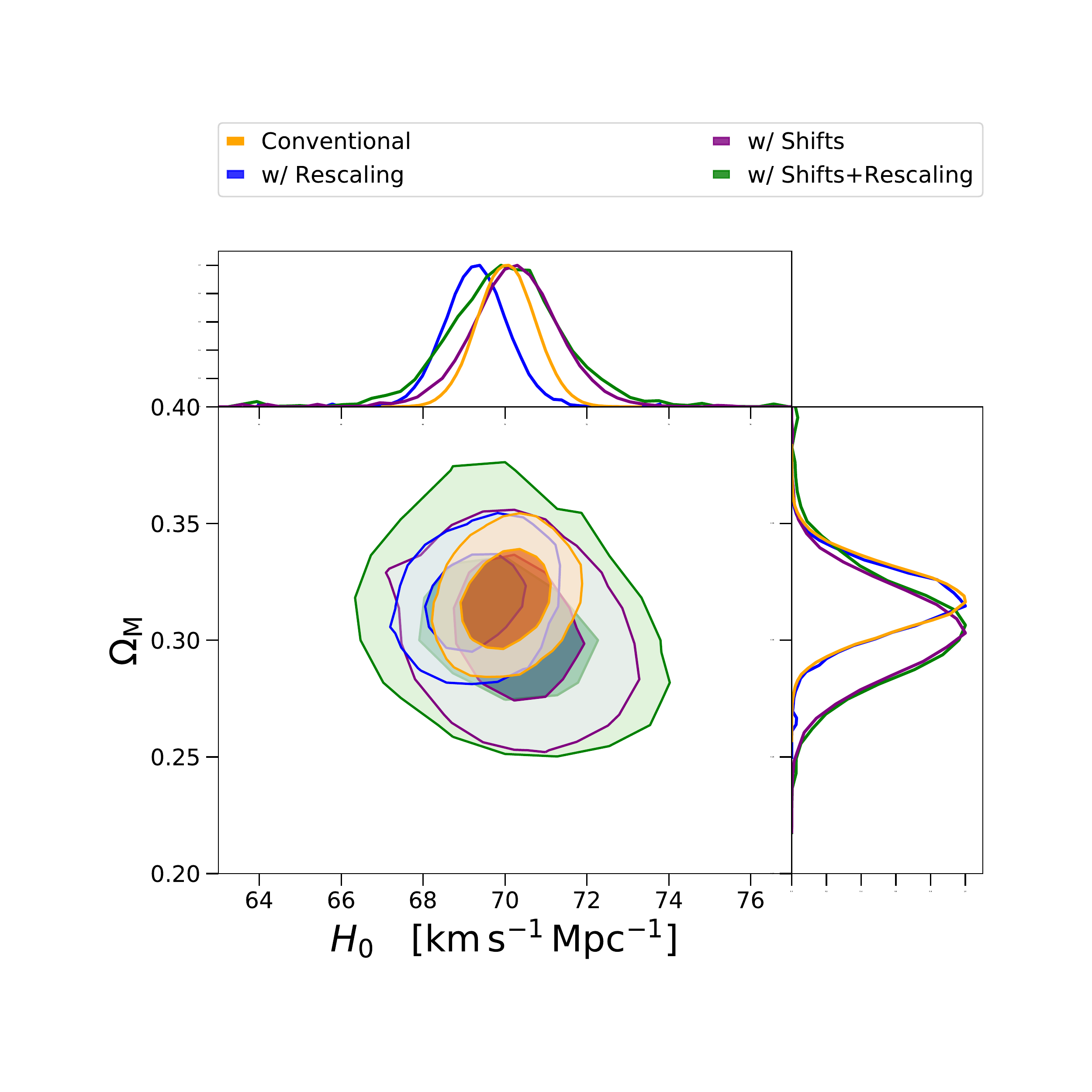}
\caption{68\% and 95\% credible level marginalized constraints on
  the $H_0$-$\Omega_{\rm M}$ plane using different methods. We show
  results with the standard approach (orange), with only rescaling
 parameters (blue), with only shift 
  parameters (purple) and
  with both rescaling and shifts 
   (green). Shifts
    are applied only to $H_0$.}
\label{fig:H0_OmegaM_onlyH0}
\end{figure}

\begin{figure}[h!]
\centering
\includegraphics[width=0.75\textwidth]{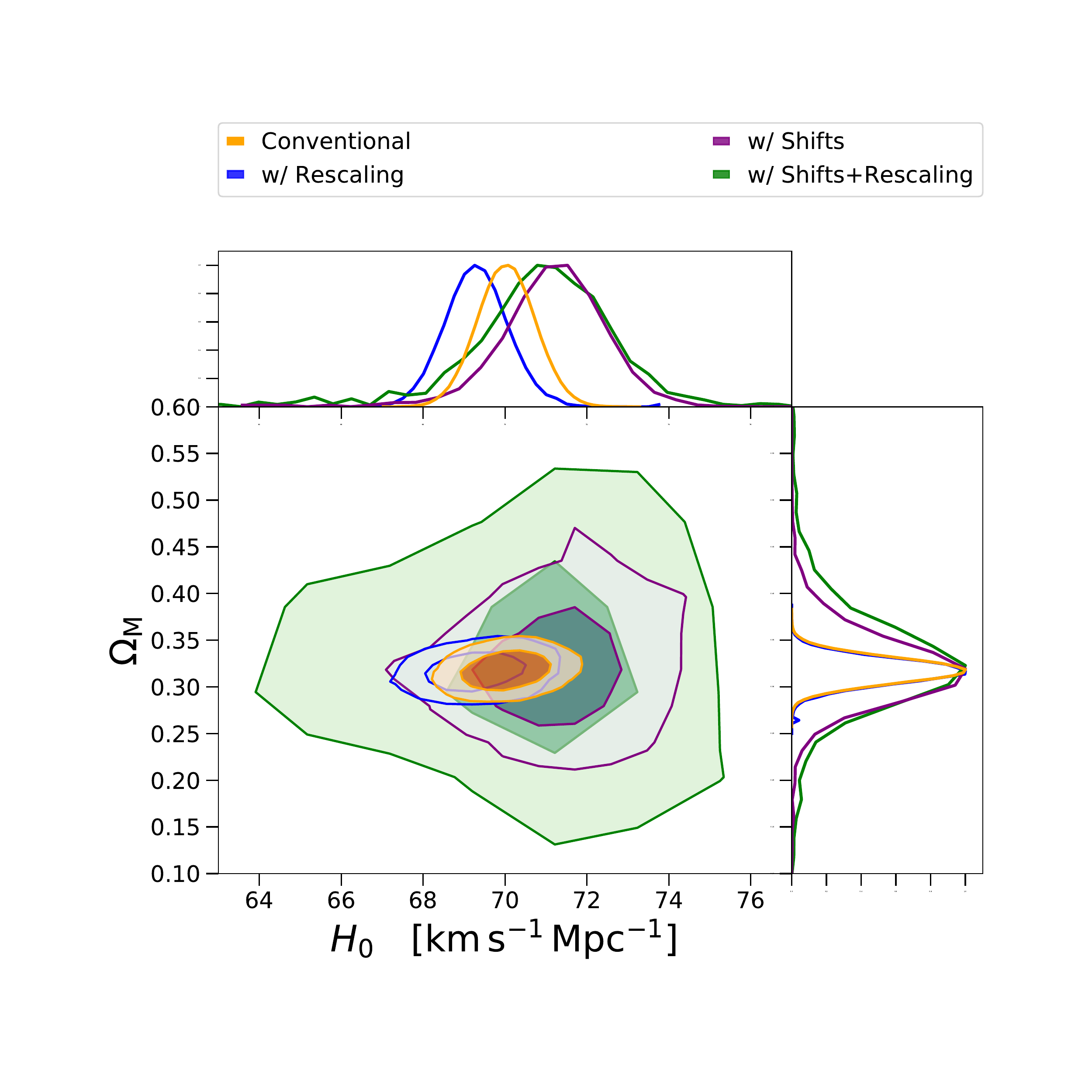}
\caption{Same as Figure \ref{fig:H0_OmegaM_onlyH0} but in this case
  the shifts 
   are applied to both $H_0$ and
  $\Omega_ch^2$. Note the change of scale in the vertical axis.}
\label{fig:H0_OmegaM}
\end{figure}

\begin{table}[htbp]
\centering
\begin{tabular}{|c|l|l|}
\hline
		Approach	&	$H_0$ ($\kmsmpc$) & \quad\quad\quad$\Omega_M$ \\
\hline
Conventional combination & \ $70.15^{+0.5}_{-0.6}(^{+1.3}_{-1.4})$ & $0.32\pm 0.01(\pm 0.03)\bigstrut$ \\
	 Rescaling param.	& \ $69.4\pm0.7(\pm 1.5)$ & $0.32\pm 0.01(\pm 0.03)\bigstrut$\\
	 Shift param.	(only $H_0$)		& \ $70.6^{+0.8}_{-1.1}(^{+ 1.9}_{-2.3})$ & $0.31\pm-0.02(\pm 0.04)\bigstrut$ \\
 Shift (only $H_0$) + rescaling param. & \ $70.5^{+0.9}_{-1.3}(^{+ 2.6}_{-3.1})$ & $0.31\pm-0.02(^{+0.04}_{-0.05})\bigstrut$ \\
 Shift param.	($H_0$ \& $\Omega_M$)		& \ $71.7^{+0.8}_{-1.2}(^{+2.0}_{-2.8})$ & $0.33\pm 0.04(^{+0.09}_{-0.07})\bigstrut$ \\
 Shift ($H_0$ \& $\Omega_M$) + rescaling param. & \ $71.0^{+1.8}_{-0.9}(^{+3.6}_{-5.4})$ & $0.33^{+0.02}_{-0.04}(^{+0.12}_{-0.14})\bigstrut$ \\
\hline

\end{tabular}
\caption{Highest posterior density values and 68$\%$ (95$\%$ in
  parenthesis) credible level marginalized constraints of $H_0$ and
  $\Omega_M$ obtained using the data and methodology described in
  Section \ref{sec:Data}.}
\label{tab:H0_OmegaM}
\end{table}

We now apply our method to the data described in
Section~\ref{sec:Data} to obtain conservative limits on $H_0$ using
all the available independent low redshift observations. 
 Regarding the introduction of shift parameters, we consider two cases. In the first case (shown
in Figure~\ref{fig:H0_OmegaM_onlyH0}) we only use them on $H_0$,
$\bDelta_{H}$. On the other hand, in the second case (shown in
Figure~\ref{fig:H0_OmegaM}) we also use them on $\Omega_ch^2$,
$\bDelta_\Omega$. 
 In both cases, rescaling 
parameters are
applied to every class of experiments and we use the same parameters
as in the previous case for the hyperprior for $\sigma_{H}$ and a
lognormal distribution with $b=-4$ and $\xi=9$ as the hyperprior for
$\sigma_\Omega$. We use $\eta=1$ for the LKJ hyperprior of the
correlation. Marginalized credible limits from both cases can be found in Table \ref{tab:H0_OmegaM}. 

As there is no inconsistency in $\Omega_M$ among the experiments
(given that most of the constraints are very weak) the only effect of
including $\bDelta_{\Omega}$ in the marginalized constraints in
$\Omega_M$ is to broaden the posteriors. In contrast, including
$\bDelta_H$ shifts the peak of the $H_0$ marginalized posterior.
While the tightest individual constraints correspond to low values of
$H_0$ (BAO and cosmic clocks), $\BACCUS$ favours slightly larger
values than the conventional approach (which stays in the middle of
the tension, as expected). These effects are larger when
we include $\bDelta_\Omega$, given that there is more freedom in the
parameter space. On the other hand, as BAO and cosmic clocks
are the largest data sets, the analysis with only rescaling
parameters prefers a lower $H_0$.  Nonetheless, as the constraints weaken when introducing shifts and rescaling, all these modifications are not of great statistical
significance.

When including only $\bDelta_H$, there is an effect on both the
constraints on $H_0$ and also on $\Omega_M$ (both slightly shifting
the maximum and broadening the errors), due to the small correlation
between the two parameters. The behaviour of the marginalized
constraints on $H_0$ is similar to the one discussed above. However,
when both $\bDelta_H$ and $\bDelta_\Omega$ are included, the
constraints are much weaker than in the previous case.  Including
shifts 
 for $\Omega_ch^2$ also increases the
uncertainties in the marginalized constraints on $H_0$.  Nonetheless,
it is important to bear in mind that the data used in this analysis
constrain $H_0$ much better than they do $\Omega_M$, even using the
conventional approach. Finally, note that in this case the constraints
 including both shift and rescaling parameters and those using only shifts 
 are not very different (in contrast to the cases
showed in Figures \ref{fig:comp_cons}, \ref{fig:comp_incons1} \&
\ref{fig:comp_incons2}), since here the type 1 systematic errors are
well accounted for and individual $\chi_\nu^2\simeq 1$.

Regarding the ability of \texttt{BACCUS} to spot which data set is more likely to be affected by systematics, there is not a clear answer for this specific problem. This is because more independent data is needed in order to discriminate between different classes of experiments, given the inconsistencies within the data sets listed in Section \ref{sec:Data}.

\section{Summary and discussion}\label{sec:Conclusions}

In this paper, we have considered the increasingly common issue of
statistical tensions in the results of cosmological experiments: small
inconsistencies in estimated parameters that are of marginal
significance, but which are too large for comfort.   
In this case, we face the
statistical question of how to combine data sets that are in tension,
in order to obtain parameter constraints that are robust. If there are
`unknown unknowns' in the the data or the theory, then the standard
analysis of the combined constraints on model parameters may not be
reliable -- which in turn risks erroneous claims of new physics in a
distinct way. 
 This is indeed a statistical issue that is not confined to cosmology:
similar challenges arise elsewhere in astrophysics
(e.g. \cite{Lee_calibXray}), and analogous challenges can be
encountered in particle physics experiments.

In response to this situation, we have introduced $\BACCUS$, a method
for combining data for parameter inference in a conservative and
agnostic way that allows consistently for the possible presence of
unknown systematics in the data sets. It deals not only with
systematics arising from incorrect estimation of the magnitude of
random measurement errors (already considered by Hobson et al. 2002;
H02), but also with those systematics whose effect is such that the
inferred model parameters are biased with respect to the true
values. The latter are the truly dangerous systematics, since they
cannot be detected by any internal null test of a single experiment.
In order to account for such effects, we introduce `shift'
parameters, $\lbrace {\bDelta}_{\btheta} \rbrace$, which offset
the best-fitting model parameters for 
each set of data independent from the rest. The magnitude of
such offsets can be constrained by inspecting the degree of agreement
between different data sets, and conservative posteriors on parameters
can be inferred by marginalizing over the offsets.

Our approach is democratic and also pessimistic: we assume that all
experiments are equally likely to suffer from shift systematics of
similar magnitude, independent of their quoted statistical precision,
and we are reluctant to set an upper limit to the size of possible
systematics.  
 Crucially, therefore,
the prior for the shifts should take no account of the size of the
reported random errors, since shift systematics by definition cannot
be diagnosed internally to an experiment, however precise it may be.
In practice, we assume that the shifts have a Gaussian distribution,
with a prior characterised by some unknown covariance matrix.  We
  adopt a separation strategy to address the hyperprior for this
  covariance, using the LKJ distribution for the correlations and
  independent lognormal distributions for the standard deviations. We
  recommend agnostic wide hyperpriors, preferring to see explicitly
how data can rein in the possibility of arbitrarily large
systematics. 

For each data set, the shift parameters are assumed to be
  drawn independently from the same prior. But this assumption is not
  valid when considering independent experiments that use the same
  technique, since they may well all suffer from systematics that are
  common to that method. Therefore data should first be combined into
  different {\it classes\/} of experiments before applying our method.
  In practice, however, a single experiment may use a number of
  methods that are substantially independent (e.g. the use of lensing
  correlations and angular clustering by DES).  In that case, our
  approach can be similarly applied to obtain conservative constraints
  and assess internal consistency of the various sub-methods.

Because it is common for joint posterior distributions to display
approximate degeneracies between some parameters, a systematic that
affects one parameter may induce an important shift in others. For
example, in Figure \ref{fig:H0_OmegaM_onlyH0} the probability density
function of $\Omega_{\rm M}$ changes due to $\bDelta_H$. For
complicated posteriors, it is therefore better in principle to use our
approach at the level of the analysis of the data (where all the model
parameters are varied), rather than constructing marginalized
constraints on a single parameter of interest and only then
considering systematics.  

These assumptions could be varied: in some cases there could be enough
evidence to consider certain experiments more reliable than others, so
that the prior for the shifts 
 will not be
universal.  But recalling the discussion in \ref{sec:Gaussian}
concerning the use of different shift priors for each data sets, a way
to proceed might be to rescale $\sigma_{\theta^j}$ only for certain
data sets (those more trusted), but then to use the same prior for all
data sets after rescaling. If we consider the data sets
$D_{i^{\prime}}$ to be more reliable than the rest, the final prior
should be
\begin{equation}
\post(\bDelta_a\lvert\sigma_a) \propto
\frac{1}{\sigma_a^{N-1}}\exp\left[-\frac{1}{2}\sum_{i\neq
    i^{\prime}}^{N}(\Delta_a^i/\sigma_a)^2\right]\frac{1}{\sigma_a/\beta}
\exp\left[-\frac{1}{2}(\Delta_a^{i^\prime}/(\sigma_a/\beta))^2\right],
\end{equation}
where we consider the case with only one parameter $a$ for clarity and
$\beta$ is a constant $>1$. 

Another possibility is to weaken the assumption that arbitrarily
bad shift systematics are possible. One can achieve this either by
imposing explicit limits so that the shifts 
 never take
values beyond the chosen bound, or by altering the prior on the shift parameters, making it narrower. 
 Although the methodology is sufficiently flexible to accommodate such
customizations, we have preferred to keep the assumptions as few and
simple as possible. As we have seen, large shifts are automatically
disfavoured as the number of concordant data sets rises, and this
seems a better way to achieve the outcome.

It is also
possible to ascertain if a single experiment is affected by atypically
large shifts, by inspecting the marginalized posteriors for the shifts
applicable to each dataset. 
 A straightforward option now is to compute
  the relative Bayesian evidence between the models with and without
  shifts, telling us how strongly we need to include them, as
  done in H02. But this procedure needs care: consider a model with
  many parameters but only one, $\theta_j$, strongly affected by type
  2 systematics. In that case, the evidence ratio will favour the
  model without shifts, those not affecting $\theta_j$ are not 
  necessary. Therefore, the ideal procedure is to check the evidence
  ratio between models with different sets of families of shifts, 
   although this is computationally demanding.

After applying our method to some simple example models and comparing
it with the scaling of reported errors as advocated by H02, we have
applied it to a real case in cosmology: the tension in $H_0$. In
general, $H_0$ values obtained in this way are larger than either
those from the conventional approach, or the combination using the approach of H02.
  However, as our conservative uncertainties are
larger there is no tension when compared with the CMB value inferred
assuming $\Lambda$CDM. We have focused on the application to parameter
inference by shifting the model parameters for each data set. However,
it is also possible to apply the same approach to each individual
measurement of a data set, in the manner that rescaling
  parameters were used by \cite{Cardona_H0}.

We may expect that the issues explored here will continue to generate
debate in the future.  Next-generation surveys will witness
improvements of an order of magnitude in precision, yielding
statistical errors that are smaller than currently known
systematics. Great efforts will be invested in refining methods for
treating these known problems, but the smaller the statistical errors
become, the more we risk falling victim to unknown systematics. In the
analysis presented here, we have shown how allowance can be made for
these, in order to yield error bounds on model parameters that are
conservative.  We can hardly claim our method to be perfect: there is
always the possibility of global errors in basic assumptions that will
be in common between apparently independent methods. 
  Even so, we have shown that realistic credibility
intervals can be much broader than the formal ones derived using standard
methods. But we would not want to end with a too pessimistic 
conclusion: the degradation of precision need not be substantial
provided we have a number of independent methods, and provided they
are in good concordance. As we
have seen, a conservative treatment will nevertheless
leave us with extended tails to the posterior, so there is
an important role to be played by pursuing a number of independent
techniques of lower formal precision. In this way, we can
obtain the best of both worlds: the accuracy of the best
experiments, and reassurance that these have not been rendered
unreliable by unknown unknowns.

 Finally, a possible criticism of
our approach is that an arms-length meta-analysis is no substitute for
the hard work of becoming deeply embedded in a given experiment to
the point where all systematics are understood and rooted out.
We would not dispute this, and do not wish our approach to be seen as encouraging lower standards of internal statistical rigour; at best, it is 
a mean of taking stock of existing results before planning the
next steps. But we believe our analysis is useful in indicating
how the community can succeed in its efforts.

\acknowledgments

We thank Licia Verde and Charles Jenkins for useful discussion during
the development of this work, and Alan Heavens, Andrew Liddle, \& Mike
Hobson for comments which help to improve this manuscript.  Funding
for this work was partially provided by the Spanish MINECO under
projects AYA2014-58747-P AEI/FEDER UE and MDM-2014-0369 of ICCUB
(Unidad de Excelencia Maria de Maeztu).  JLB is supported by the
Spanish MINECO under grant BES-2015-071307, co-funded by the ESF. JLB
thanks the Royal Observatory Edinburgh at the University of Edinburgh
for hospitality. JAP is supported by the European Research Council,
under grant no. 670193 (the COSFORM project).

\bibliography{biblio}
\bibliographystyle{utcaps}

\end{document}